\documentclass[aps,pra,reprint, superscriptaddress,amsmath,showpacs,longbibliography]{revtex4-2}
\usepackage{xcolor}
\usepackage{tcolorbox}
	
\usepackage{graphicx}

\graphicspath{{./Figures/}}

\usepackage[colorinlistoftodos,prependcaption ]{todonotes}

\usepackage{xargs}
\newcommandx{\greencom}[2][1=]
{\todo[inline, color=green!40,#1]{#2}}
\newcommandx{\bluecom}[2][1=]
{\todo[inline, color=blue!40,#1]{#2}}
\newcommandx{\bluemargin}[2][1=]
{\todo[color=blue!40,#1]{#2}}

\newcommand*{\colorboxed}{}
\def\colorboxed#1#{%
	\colorboxedAux{#1}%
}
\newcommand*{\colorboxedAux}[3]{%
	\begingroup
	\colorlet{cb@saved}{.}%
	\color#1{#2}%
	\boxed{%
		\color{cb@saved}%
		#3%
	}%
	\endgroup
}

\usepackage{pifont}

\usepackage{float}
\usepackage{epsfig}
\usepackage{bm}
\usepackage[matrix,frame,arrow]{xy}
\usepackage[applemac]{inputenc}
\usepackage[T1]{fontenc}
\usepackage{lmodern}
\usepackage[english]{babel}
\usepackage{amsmath}
\usepackage{ae}
\usepackage{amssymb}
\usepackage{color}
\usepackage{graphicx}
\usepackage{bbm}
\usepackage{url}
\usepackage{nomencl}
\usepackage{subfigure}
\usepackage{slashed}

\usepackage{booktabs}

\usepackage[markup=underlined]{changes}
\makeatletter
\@namedef{Changes@AuthorColor}{magenta}
\colorlet{Changes@Color}{magenta}
\makeatother
\newcommand{\ket}[1]{|#1\rangle}

\newcommand{\figref}[1]{\mbox{Fig.~\ref{#1}}}

\renewcommand{\eqref}[1]{\mbox{Eq.~(\ref{#1})}}

\newcommand{\be}{\begin{equation}}
\newcommand{\ee}{\end{equation}}
\newcommand{\bea}{\begin{eqnarray}}
\newcommand{\eea}{\end{eqnarray}}
\newcommand{\beal}{\begin{align}}
\newcommand{\eeal}{\end{align}}

\newcommand{\qum}[1]{``#1''}

\usepackage{xr}
\usepackage[colorlinks]{hyperref}
\hypersetup{%
    plainpages=true,
    breaklinks=true,
    hypertexnames=false,
    pageanchor=true,
    colorlinks=true,
    linkcolor={blue},
    citecolor={red},
    urlcolor={blue},
    anchorcolor={black}
}

\makeatletter

\newcommand{\beq}{\begin{eqnarray}}
\newcommand{\eeq}{\end{eqnarray}}

\begin{document}

	\title{Gauge Principle and Gauge Invariance in Two-Level Systems}

\author{Salvatore Savasta}
\affiliation{Dipartimento di Scienze Matematiche e Informatiche, Scienze Fisiche e  Scienze della Terra,
	Universit\`{a} di Messina, I-98166 Messina, Italy}

\author{Omar Di Stefano}
\email[corresponding author: ]{odistefano@unime.it}
\affiliation{Dipartimento di Scienze Matematiche e Informatiche, Scienze Fisiche e  Scienze della Terra, Universit\`{a} di Messina, I-98166 Messina, Italy}	

\author{Alessio Settineri}
	\affiliation{Dipartimento di Scienze Matematiche e Informatiche, Scienze Fisiche e  Scienze della Terra, Universit\`{a} di Messina, I-98166 Messina, Italy}
	
\author{David Zueco}
	\affiliation {Instituto de Ciencia de Materiales de
		Arag\'{o}n and Departamento de F\'{i}sica de la Materia Condensada ,
		CSIC-Universidad de Zaragoza, Pedro Cerbuna 12, 50009 Zaragoza,
		Spain}
	\affiliation{Fundaci\'{o}n ARAID, Campus R\'{i}o Ebro, 50018 Zaragoza, Spain}
	\author{Stephen Hughes}
	\affiliation {Department of Physics, Engineering Physics, and Astronomy,
		Queen's University, Kingston, Ontario K7L 3N6, Canada}

%
%

	\author{Franco Nori}
\affiliation{Theoretical Quantum Physics Laboratory, RIKEN Cluster for Pioneering Research, Wako-shi, Saitama 351-0198, Japan} \affiliation{Physics Department, The University
	of Michigan, Ann Arbor, Michigan 48109-1040, USA}
%

%


\begin{abstract}



The quantum Rabi model is a widespread description of the coupling between a two-level system
and a quantized single mode of an electromagnetic resonator. Issues about this model's gauge invariance have been raised. These
issues become evident when the light-matter interaction reaches the so-called ultrastrong coupling regime. Recently, a modified quantum Rabi model able to provide gauge-invariant physical results (e.g.,  energy levels, expectation values of observables, quantum probabilities) in any interaction regime
{was introduced} [Nature Physics
{\bf 15}, 803 (2019)]. Here we provide an alternative derivation of this result, based on the implementation in two-state systems of the  gauge principle, which is the principle from which all the fundamental interactions in quantum field theory are derived.
The adopted procedure can be regarded as the two-site version of the general method used to implement the gauge principle in lattice gauge theories.
Applying this method, we also obtain the gauge-invariant quantum Rabi model for asymmetric two-state systems, and the multi-mode gauge-invariant quantum Rabi model beyond the dipole approximation.

\end{abstract}

	\maketitle

\section{Introduction}
Recently, it has been argued that truncations of the atomic Hilbert space, to obtain a two-level description of the matter system, violate the gauge principle \cite{DeBernardis2018, Stokes2019,Stokes2020a}. Such violations become particularly
relevant in the  ultrastrong  and deep-strong coupling (USC and DSC) regimes.
These extreme regimes have been realized between individual or collections of effective two-level systems (TLSs) and the electromagnetic field 
in a variety of settings \cite{Kockum2018, Forn-Diaz2018}. In the USC (DSC) regime of quantum light-matter interaction
the coupling strength becomes comparable   (larger) than the transition frequencies of the system.

Reference {\em et al.}~\cite{DeBernardis2018} demonstrated that,  while in the electric dipole gauge, the two-level approximation can be performed as long as the Rabi frequency remains much smaller than the energies of all higher-lying levels. The two-level approximation  can drastically
fail in the Coulomb gauge, even for systems with an extremely anharmonic spectrum.

The impact of the truncation of the Hilbert space of the matter system to only two states was  also studied ~\cite{Stokes2019}, by introducing a one-parameter ($\alpha$) set of gauge transformations. 
The authors {found} that each  value of the
parameter produces a distinct quantum Rabi model (QRM), thus providing distinct physical predictions. {I}nvestigating a {\em matter} system with a lower anharmonicity (with respect to that considered in Ref.~\cite{DeBernardis2018}),
they use the gauge parameter $\alpha$  as a  fit parameter to determine the optimal QRM for a specific set of system parameters, by comparing the obtained $\alpha$-dependent lowest energy states and levels with the corresponding predictions of the  non-truncated gauge invariant model. The surprising result~\cite{Stokes2019} is that, according to this procedure, in several circumstances the {\em optimal} gauge is the so-called Jaynes-Cummings (JC) gauge, a gauge where the counter-rotating terms are automatically absent.

Recently, the source of gauge violation has been identified {~\cite{DiStefano2019}}, and a general method for the derivation of light-matter Hamiltonians
in truncated Hilbert spaces able to produce gauge-invariant physical results has been developed{~\cite{DiStefano2019}}
(see also related work \cite{Settineri2020, Garziano2020,SDN2020}). 
This gauge invariance {was} achieved by {\em compensating the non-localities} introduced in the construction of the effective Hamiltonians. Consequently, the resulting
quantum Rabi Hamiltonian in the Coulomb gauge differs significantly  from the standard one, but provides exactly the {\em same}
energy levels obtained by using the dipole gauge, as it should, because physical observable quantities must be gauge invariant. A recent overview of these gauge issues in TLSs can be found
in Ref.~\cite{LeBoite2020}.

Very recently, the validity of the gauge invariant {QRM} developed in Ref.~\cite{DiStefano2019} has been put into question~\cite{Stokes2020a}.
Specifically, it is claimed that  the truncation of the Hilbert space necessarily ruins gauge-invariance.

In this paper, however, we confirm that the gauge principle applies also to TLSs, as required by any consistent description of light-matter interaction.
Specifically, we formulate in a fully consistent and physically meaningful way the fundamental gauge principle in two-state systems.
The derivation described here  can be regarded as the two-site version of the general method for {\em lattice gauge theories} \cite{Wiese2013}. These represent the most advanced and {commonly} used tool for describing gauge theories in the presence of {a truncated infinite-dimensional Hilbert space}. When a gauge theory is regularized on the lattice, it is vital to maintain its invariance
under gauge transformations~\cite{Wiese2013}. An analogous approach has been developed as early as 1933~\cite{Peirls1933} for the description of  tightly-bound electrons in a crystal in the presence of a slowly-varying magnetic vector potential (see, e.g., also {Refs.}~\cite{Luttinger1951, Hofstadter1976, Graf1995}).
Applying this method, we also obtain the multi-mode gauge-invariant {QRM} beyond the dipole approximation.

\section{The gauge principle}

In this section, we recall some fundamental concepts, which we will apply in the next sections.

In quantum field theory, the coupling of particles with fields is constructed in such a way
that the theory is invariant under a gauge transformation{~\cite{Maggiore2005}}. Here, we limit the {theoretical} model to consider  $U(1)$ invariance.
For symmetry groups that are non-commutative,
this approach can be generalized to 
non-abelian gauge theories \cite{Maggiore2005, Wiese2013}.

Let us consider the transformation of the particle field $\psi \to \exp(i q \theta) \psi$.
This transformation represents a symmetry of the free action of the particle (e.g., the Dirac action) if $\theta$ is a constant, but we want to consider a generic function $\theta(x)$
({\it local} phase transformation).
However, the free Dirac action is not invariant under local phase transformations, because the factor  $\exp[i q \theta(x)]$ does not commute with $\partial_\mu$.
At the same time, it is known that the action of the free electromagnetic field is invariant
under the 
following gauge transformation:
\be\label{gaugeA}
A_\mu \to A_\mu - \partial_\mu {\theta} \, .
\ee
It is then possible to replace, in the action, the derivative $\partial_\mu$ with a {\em covariant derivative} of $\psi$ as
\be 
D_\mu \psi = \left( \partial_\mu + i q A_\mu \right) \psi\, ,
\ee
so that
\be
D_\mu \psi \to e^{i q \theta} D_\mu \psi\, ,
\ee
even when $\theta$ depends on $x$. 
It is now easy to construct a Lagrangian with a local $U(1)$ invariance.
It suffices to replace all derivatives $\partial_\mu$ with covariant derivatives $D_\mu$.

The same procedure, leading to the well-known minimal coupling replacement, can be applied to describe the interaction of a non-relativistic particle with the electromagnetic field. 
Considering a particle of mass {$m$} with a geometrical coordinate $x$ and a potential $V(x)$, the Hamiltonian of such a particle interacting with the electromagnetic field can be written as
\be\label{mcr}
\hat H_0^{\rm gi} = \frac{1}{2 m} \left[\hat p - q A (x) \right]^2  + V(x)\, ,
\ee
where $\hat p = - i d/dx$ is the momentum of the particle (here $\hbar =1$).
It turns out that the expectation values $\langle \psi | \hat H_0^{\rm gi} | \psi \rangle$ are invariant under local phase transformations,
\be\label{phasex}
\psi(x) \to e^{iq \theta(x)} \psi (x)\, ,
\ee
 thanks to the presence of the gauge field $A(x)$.

Note that the function of a continuous degree of freedom $\psi (x)$ lives in the infinite-dimensional space of all square-integrable functions, and the local phase transformation transforms a state vector in this space into a different vector in the same space. 
Finally, we observe that the total Hamiltonian, in addition to $\hat H_0^{\rm gi}$, includes the free Hamiltonian for the gauge field.

\section{Double-well systems  in the two-state limit}\label{doublewell}
\begin{figure}[htpb] 
	\centering
	\includegraphics[width= 0.75 \linewidth]{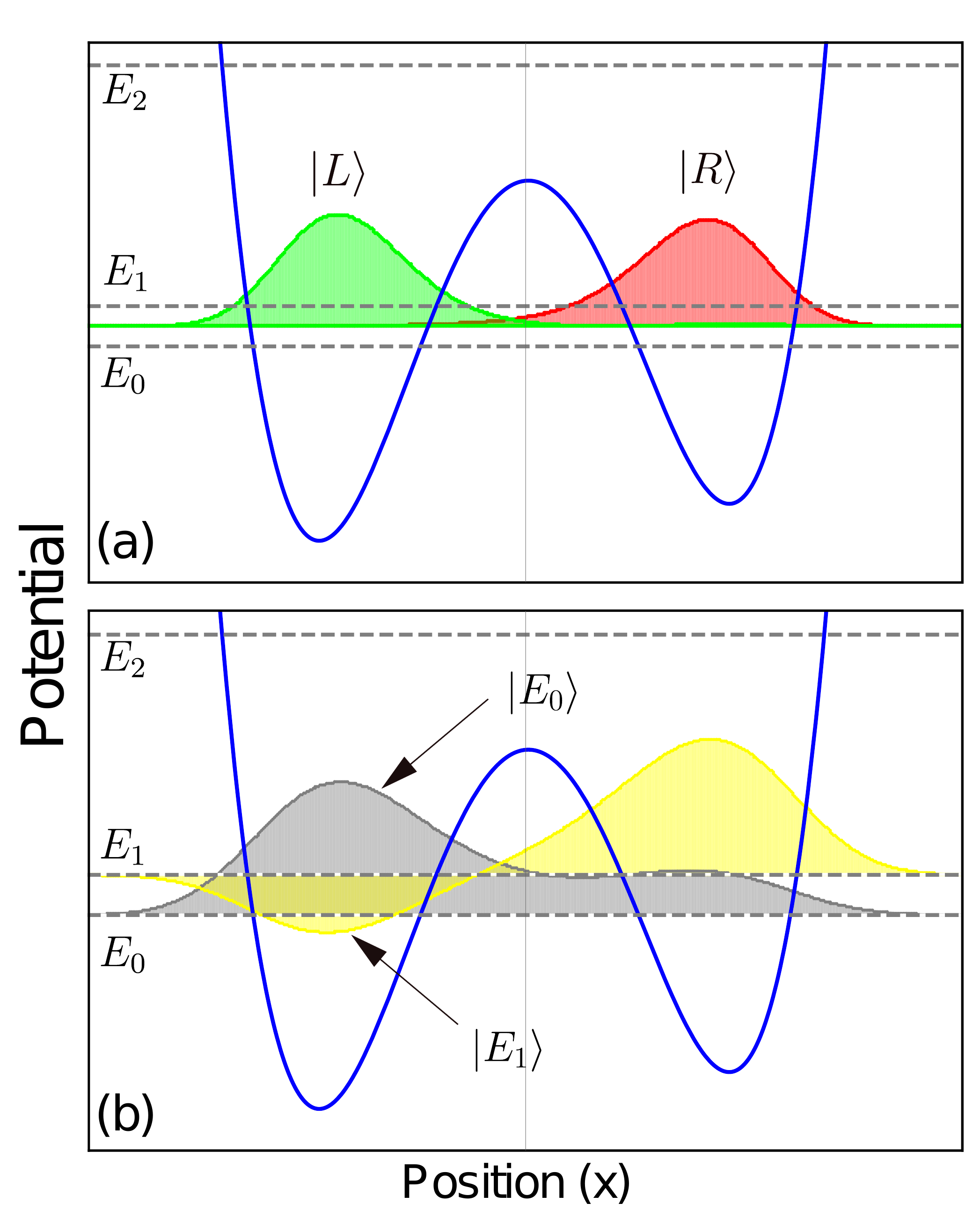} 
	\caption{{\bf A double-well system in the {\em two-state} limit}. {The symbols} $E_0$ and $E_1$ are the two lowest-energy levels, well separated in energy by the next higher energy level $E_2$. Panel (a) also shows the square modulus of the two wavefunctions localized in the well, obtained as linear combinations of the two lowest energy wavefunctions displayed in panel (b).}

	\label{fig1}
\end{figure}

The problem of a quantum-mechanical system whose
state is effectively restricted to a two-dimensional Hilbert
space is ubiquitous in physics and chemistry \cite{Leggett1987}. In the simplest
examples, the system 
simply possesses a degree of
freedom that can take only two values. For example, the
spin projection in the case of a 
spin-$1/2$ particle  or the polarization
in the case of a photon. Besides these
{\em intrinsically} two-state systems, a more common situation
is that the system has a continuous
degree of freedom $x$, for example, a geometrical coordinate, and a potential energy function
$V(x)$ depending on it, with two separate minima \cite{Leggett1987} (see \figref{fig1}). 
 Let us assume that the barrier height $V$ is large enough that the system dynamics can be adequately described by a two-dimensional Hilbert space spanned by the two {\em ground} states in the two wells $| L \rangle$ and $| R \rangle$.

The motion in the two-dimensional Hilbert space can be adequately described by the simple Hamiltonian:
\be \label{H01}
\hat {\cal H}_0 =   \sum_{j= L,R} E_j | j \rangle \langle j  | - t \left( |R \rangle \langle L |+ {\rm h. c.}  \right)\, ,
\ee
where the tunneling coefficient is given by $t = \langle L | \hat H_0 | R \rangle$, and
\be\label{H0f}
\hat H_0 = \frac{\hat p^2}{2m} + V(x) 
\ee
is the usual system Hamiltonian.

\begin{figure}[htpb]  
	\centering
	\includegraphics[width= 0.75 \linewidth]{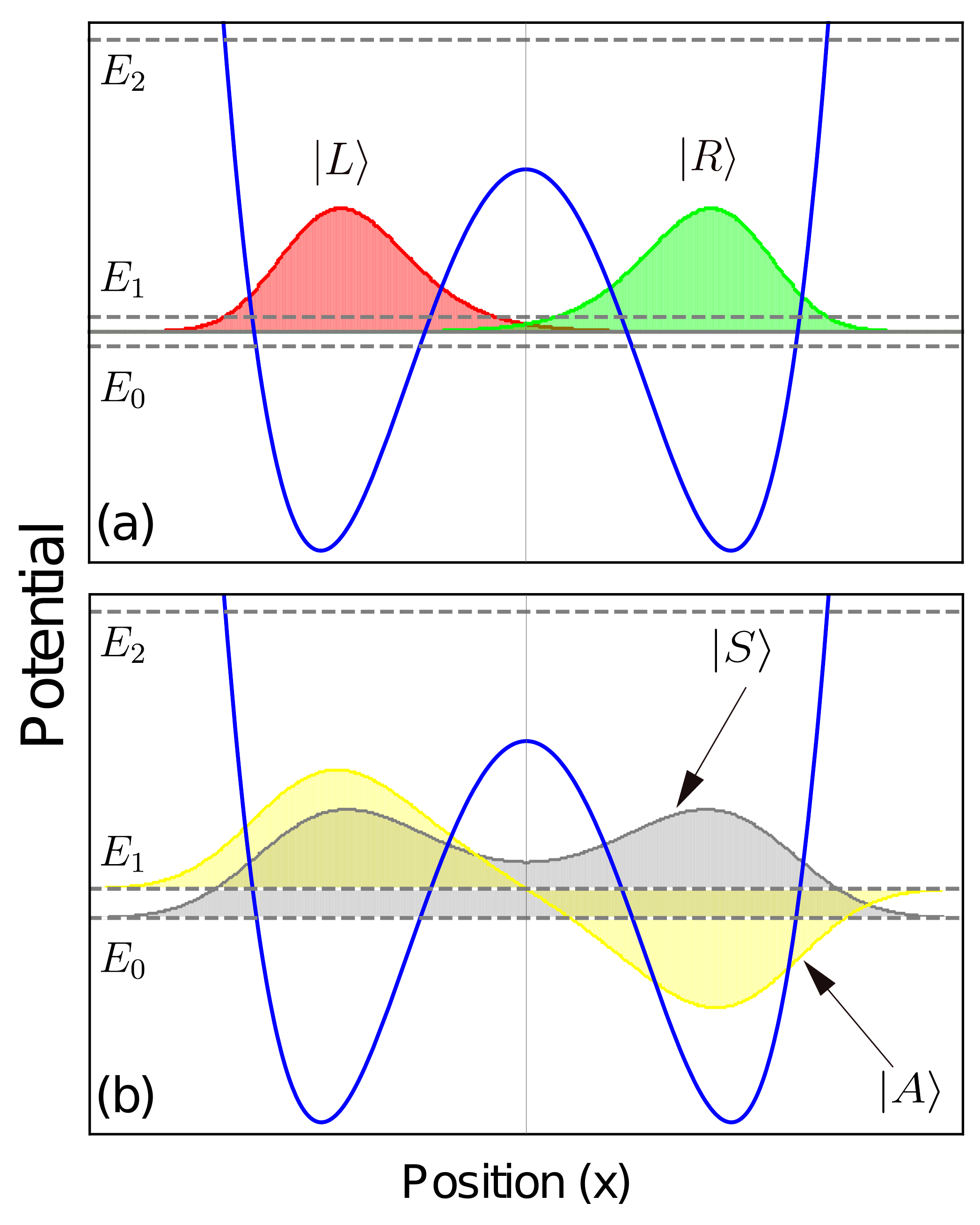}  
	\caption{{\bf A symmetric double-well system in the {\em two-state} limit}. {The symbols} $E_0$ and $E_1$ are the two lowest-energy levels, well separated in energy by the next higher energy level $E_2$. Panel (a) also shows the square modulus of the two wavefunctions localized in the well, obtained as symmetric and antysimmetric combinations of the two lowest energy wavefunctions displayed in panel (b).}

	\label{fig2}
\end{figure}

If the potential is an even function of the geometrical coordinate, namely $V(x) = V(-x)$ (see \figref{fig2}), then  $E_L = E_R$, and we can fix $E_L = E_R = 0$. Introducing the  Pauli operator 
 $\hat \rho_x = |L \rangle \langle R |+ {\rm h.c.}$, we {obtain}
\be\label{H02}
\hat {\cal H}_0 = -t \hat \rho_x\, ,
\ee
whose eigenstates, delocalized in the two wells, are the well-known symmetric- and antisymmetric combinations (see \figref{fig2}b),
\bea\label{SA}
| S  \rangle &=& \frac{1}{\sqrt{2}} \left( |R \rangle + | L \rangle \right)\, , \nonumber \\
| A \rangle &=& \frac{1}{\sqrt{2}} \left (|R \rangle - |L\rangle \right)\, ,
\eea
with eigenvalues $E_{A,S} = \pm t$, so that $\Delta = E_A - E_S = 2t$, where we {assume} $t>0$.
The Hamiltonian in \eqref{H01} can be written in diagonal form as
\be \hat {\cal H}_0 = (\Delta/2) \hat \sigma_z\, ,
\ee
where $\hat \sigma_z = - \hat \rho_x = |A \rangle \langle A| - |S \rangle \langle S|$.
{Note, to distinguish between the different basis states for the operator representations,
we  use $\hat \sigma_i$
for the $\ket{A}{-}\ket{S}$ basis,
and $\hat \rho_i$ for the
$\ket{L}{-}\ket{R}$ basis.
Thus, for example, the diagonal $\hat \sigma_z$
operator becomes nondiagonal
in the $\ket{L}{-}\ket{R}$ basis.
}

It is worth noting that this elementary analysis is not restricted to the case of a double-well potential. Analogous considerations can be carried out for systems with different potential shapes, displaying two (e.g., lowest energy) levels well separated in energy from the next higher level.
The wavefunctions $\psi_L(x) = \langle x |L \rangle$ and $\psi_R(x) = \langle x |L \rangle$ can be obtained from the symmetric and antisymmetric combinations of $\psi_S(x)$ and $\psi_A(x)$ (see \figref{fig1}), which can be obtained exactly as the two lowest energy eigenfunctions of the  Schr\"{o}dinger problem described by the Hamiltonian in \eqref{H0f}. The gap $\Delta = 2 t$ is obtained from the difference between the corresponding eigenvalues.
This two-state tunneling model is a well known formalism
to describe many realistic systems,
including the ammonia molecule, coupled quantum dots, and superconducting flux-qubits.

The case of a potential of the effective particle which does not display inversion symmetry can also be easily addressed. For example, consider an asymmetric double well potential, as shown in \figref{fig1}.  
In this case,  \eqref{H01} can be expressed as 
\be\label{H0a1}
\hat {\cal H}_0 = \frac{\epsilon}{2} \hat \rho_z - \frac{\Delta}{2} \hat \rho_x\, .
\ee
The quantity $\epsilon$ is the {\em detuning} parameter,
that is, the difference in the ground-state energies of the states localized in the two wells in the absence of tunneling. The Hamiltonian in \eqref{H0a1} can be trivially diagonalized with eigenvalues $\pm \omega_q/2$, where $\omega_q = \sqrt{\Delta^2 + \epsilon^2}$.

\section{The gauge principle in two-level systems}
\label{GaugeTLS}

The question arises if it is possible to {\em save the gauge principle} when, under the conditions described above, such a particle is adequately described by states confined {\em in a two-dimensional complex space}.
If we apply an arbitrary local phase transformation to, e.g., the wavefunction $\psi_A(x) = \langle x | A \rangle$: $\psi_A(x) \to \psi'_A(x) = e^{i q \theta(x) }\psi_A(x)$, it happens that, in general, $\psi'_A(x) \neq c_S \psi_S(x) + c_A \psi_A(x)$, where $c_A$ and $c_S$ are complex coefficients. Thus the general local phase transformation does not guarantee that the system can still be described as a two state system. According to this analysis, those works claiming {\em gauge non-invariance due to material truncation in ultrastrong-coupling QED}
\cite{Stokes2019a} (we would say at any coupling strength, except negligible),  at first sight, might appear to be correct. The
direct consequence of this conclusion would be that two-level models, widespread in physics and chemistry, are too simple to implement their interaction with a gauge field, according to the general principle from which the fundamental interactions in physics are obtained. Since adding to the particle system description a few additional levels does  not change this point, the conclusion would be even more dramatic. Moreover, according to Ref.s~\cite{Stokes2019, Stokes2020a},  this leads to several non-equivalent models of light-matter interactions {\em providing different physical results}. 
One might then naively claim the \qum{death of the gauge principle} and of gauge invariance in truncated Hilbert spaces, 
namely in almost all cases where theoreticians try to provide quantitative predictions to be compared with actual experiments. 

Our view is drastically different: we find that the breakdown of gauge invariance is the direct consequence of the {\em inconsistent} approach  of reducing the information (Hilbert space truncation) on the effective particle,  without accordingly reducing the information, by the same amount, on the phase $\theta(x)$ determining the transformation in \eqref{phasex}. In physics, the approximations must be done with care, and they must be consistent.

We start by observing that the two-state system defined in \eqref{H01} {still has} {\em a geometric coordinate}, which however {\em can assume only two values}: $x_j$ (with $j= L, R$), {that} we can approximately identify with the position of the two minima of the double-well potential. More precisely, and {more} generally, they are:
\bea\label{xr}
x_R &=&  \langle R | x | R \rangle
, \nonumber \\
x_L &=& \langle L | x | L \rangle\, .
\eea
Here, parity symmetry implies $x_L = - x_R$. In the following we will use the shorthand $\langle R | x | R \rangle = a/2$. Hence, the operator describing the geometric coordinate can be written as~\cite{Leggett1987}
${\cal X} = (a/2) \hat \rho_z$, where $\hat \rho_z \equiv | R \rangle \langle R| - | L \rangle \langle L |$.

We observe that the terms proportional to $t$ in the Hamiltonian in \eqref{H01} or \eqref{H02}, implies that these can be regarded as {\em nonlocal} Hamiltonians, i.e., with an effective potential depending on two distinct coordinates. Nonlocality here comes from the hopping term  $t = \langle R | \hat H_0 | L \rangle$, which is determined by the interplay of the kinetic energy term and of the potential energy in $\hat H_0$.

It is clear that the consistent and meaningful local gauge transformation corresponds to the following transformation:
\be\label{phasec}
| \psi \rangle = c_L |L \rangle + c_R | R \rangle 
\to | \psi' \rangle = e^{i q \theta_L} c_L |L \rangle + e^{i q \theta_R}  c_R | R \rangle\, ,
\ee
where $| \psi \rangle$ is a generic state in the two-dimensional Hilbert space, and $\theta_j$ are arbitrary real valued parameters. 

It is easy to show that the expectation values of $\hat {\cal H}_0$ are not invariant under the {\em local } transformation in \eqref{phasec}.
They are only invariant under a uniform phase change: $| \psi \rangle \to \exp(i q \theta)| \psi \rangle$.
 However, one can introduce in the Hamiltonian field-dependent factors, that {\em compensate the difference in the phase transformation from one point to the other}. Specifically, following the general procedure of lattice gauge theory, we can {consider} the {\em parallel transporter} (a unitary {\it finite-dimensional} matrix), introduced by Kenneth Wilson \cite{Wilson1974,Lang2010,Wiese2013}, 
\be\label{W1}
U_{x_k +a,x_k} = \exp{\left[ i q \int_{x_k}^{x_k+a} dx\, A(x)   \right]}\, ,
\ee 
where $A(x)$ is the gauge field. After the gauge {transformation} of the field, 
\be
A'(x) = A(x) + d \theta / dx\, ,
\ee
the transporter transforms as
\be\label{W2}
U'_{x_k +a,x_k} = e^{i q\, \theta(x_k+a)} U_{x_k +a,x_k} e^{-i q\, \theta(x_k)}\, 
\ee
which is now discrete.
This property can  also be used to implement gauge invariant Hamiltonians in two-state systems.

\subsection{Symmetric two-state systems}

Properly introducing the parallel transporter in \eqref{W1} into \eqref{H02}, we obtain a gauge-invariant two-level model:
\be\label{HL1}
\hat {\cal H}^{\rm gi}_0 = - t\, |R \rangle \langle L|\, U_{x_R, x_L} + {\rm h. c.}\, .
\ee
Gauge invariance can be directly verified: 
\bea
&\langle& \psi' | \left( |R \rangle \langle L|\, U'_{x_R, x_L} + {\rm h. c.} \right) |\phi' \rangle =  \nonumber \\
&\langle& \psi | \left( |R \rangle \langle L|\, U_{x_R, x_L} + {\rm h. c.} \right) |\phi \rangle\, ,\nonumber
\eea
where $| \psi \rangle$ and $| \phi \rangle$ are two generic states in the vector space spanned by $| L \rangle$ and $| R \rangle$.
By neglecting the spatial variations of the field potential $A(x)$ on the distance $$a = x_R - x_L\, ,$$ (dipole approximation). The Hamiltonian in \eqref{HL1} can be written as
\be\label{HLa}
\hat {\cal H}^{\rm gi}_0 = - t\, |R \rangle \langle L|\, e^{i q a A} + {\rm h. c.}\, .
\ee
Using \eqref{SA} and the Euler formula, it can be easily verified that the Hamiltonian in \eqref{HLa} can be expressed using the diagonal basis of $\hat {\cal H}_0$, as
\be\label{HL2}
\hat {\cal H}^{\rm gi}_0 =  \frac{\Delta}{2} \left[ \hat \sigma_z
 \cos{(q a  A)} + \hat \sigma_y \sin{(q a  A)}
 \right]\, ,
\ee
where $\hat \sigma_y = -i \left( | A \rangle \langle S| - |S \rangle \langle A| \right)$. Using \eqref{SA} and \eqref{xr},  {then} 
\be\label{d}
q a/2 = q \langle A | x | S \rangle\, . 
\ee
This {precisely} coincides with the transition matrix element of the dipole moment as in Ref. \cite{DiStefano2019}. 

Considering a quantized field $\hat A$, the total light-matter Hamiltonian also contains the free-field contribution, $\hat H_{\rm ph}$, so that:
\be\label{HL3}
	\hat {\cal H} =  \frac{\Delta}{2} \left[ \hat \sigma_z
	\cos{(q a \hat A)} + \hat \sigma_y \sin{(q a \hat A)}
	\right] 
	+ \hat H_{\rm ph}\, .
\ee
For the simplest case of a single-mode electromagnetic resonator,  the potential can be expanded in terms of the mode photon destruction and creation operators. Around $x=0$,  $\hat A = A_0 (\hat a + \hat a^\dag)$, where $A_0$ {(assumed real)} is the zero-point-fluctuation amplitude of the field in the spatial region spanned by the effective particle. We also have: $\hat H_{\rm ph} = \omega_{\rm ph} \hat a^\dag \hat a$, where $\omega_{\rm ph}$ is the resonance frequency of the cavity mode. It can be useful to define the normalized coupling strength parameter \cite{DiStefano2019}
\be
\eta = q (a/2) A_0\, ,
\ee
so that \eqref{HL3} can be written as
\bea
\hat {\cal H} &=&  \frac{\Delta}{2} \left\{ \hat \sigma_z
\cos{[ 2 \eta (\hat a + \hat a^\dag) ]} + \hat \sigma_y \sin{[2 \eta (\hat a + \hat a^\dag)]}
\right\}\nonumber \\
&+& \omega_{\rm ph} \hat a^\dag \hat a\, .
\label{HL33}
\eea

Using  the relations $\hat \rho_z \equiv | R \rangle \langle R| - |L \rangle \langle L| = |A \rangle \langle S| + |S \rangle \langle A|\equiv \hat \sigma_x$,
the Hamiltonian in \eqref{HL1} can also be expressed as
\be\label{HL4}
\hat{\cal H} = \hat {\cal U} \hat {\cal H}_0 \hat {\cal U}^\dag\, ,
\ee
where 
\be\label{U0}
 \hat {\cal U} = \exp{(i q a \hat A \hat \sigma_x/2)}\, .
\ee
Equations (\ref{HL4}) and (\ref{U0}) coincide with Eqs. (8) and (9) of Ref.~\cite{DiStefano2019}, which represents 
our main results. 

It is also interesting to rewrite the {\em coordinate}-dependent phase transformation in \eqref{phasec} as the application of a unitary operator on the system states. Defining $\phi = (\theta_R + \theta_L)/2$ and $\theta = (\theta_R - \theta_L)/2$, \eqref{phasec} can be written as 
\be\label{phasec2}
| \psi \rangle \to | \psi' \rangle = e^{i q \phi} e^{i q \theta \hat \sigma_x} | \psi \rangle\, .
\ee
This shows that the {\em coordinate}-dependent phase change of a generic state of a {TLS} is equivalent to a global phase change, which produces no effect, plus {\em a rotation in the Bloch sphere}, which can be compensated by  introducing a gauge field as in \eqref{HL4}.
Notice {also} that {\eqref{phasec2} coincides} with the result presented in the first section of the  {Supplementary Information of Ref.~\cite{DiStefano2019}}, obtained with a different, but equivalent approach.
\vspace{0.4 cm}

In summary, the method {presented here} can be regarded as the two-site version (with the additional dipole approximation) of the general method for {\em lattice gauge theories} \cite{Wiese2013}, which represents the most advanced and {sophisticated} tool for describing gauge theories in the presence of truncation of infinite-dimensional Hilbert spaces.
These results eliminate any concern about the validity of the results presented in Ref.~\cite{DiStefano2019}, raised 
by Ref.~\cite{Stokes2020a}.

We conclude this subsection by {noting} that \eqref{HL1} can be also used, without applying the dipole approximation, to obtain  the (multi-mode) {\em gauge-invariant quantum Rabi model beyond the dipole approximation}.	
Specifically, without applying the dipole approximation to \eqref{HL1}, after the same steps to obtain \eqref{HL33}, we obtain
\bea
	\hat {\cal H} &=&  \frac{\Delta}{2} \left[ \hat \sigma_z
	\cos\!{\left(q\! \int^{x_R}_{x_L}\! dx\, \hat A(x)\right)} \right. \nonumber \\ &+& \left. \hat \sigma_y \sin\!{\left(q\! \int^{x_R}_{x_L}\! dx\, \hat A(x)\right)}
	\right] 
	+ \hat H_{\rm ph}\, .
\label{Hbd}
\eea
One interesting consequence of this result is that it 
{introduces} {\em a natural cut-off} for the interaction of high energy modes of the electromagnetic field with a TLS. In particular, owing to cancellation effects in the integrals in \eqref{Hbd}, the resulting coupling strength between the TLS and the mode goes rapidly to zero when the mode wavelength becomes shorter than $a/2 = \langle A | x | S \rangle$.
This finding can stimulate further investigations beyond the dipole approximation, without having to introduce a cut-off frequency by hand.

It is worth noticing that  this derivation of the gauge-invariant QRM does not require the introduction of an externally controlled  two-site  {\em lattice} spacing, in contrast to general lattice gauge theories. In the present case, the effective spacing $a$ between the two sites is only determined by the transition matrix element of the position operator between the two lowest energy states of the effective particle, $a = 2 \langle A | x | S \rangle$, which in turn determines the dipole moment of the transition, $q a /2$.

\subsection{Asymmetric two-state systems}
The results in this section can be directly generalized to also address the case of a potential of the effective particle which does not display inversion symmetry.
It has been shown that the interaction (in the USC and DSC limit) of these TLSs (without inversion symmetry), with photons in resonators, can lead to a number of  interesting phenomena \cite{Niemczyk2010, Ridolfo2012, Garziano2015, Garziano2016,Yoshihara2017,Kockum2017a, Stassi2017}.
In this
case, \eqref{H0a1} provides the bare TLS Hamiltonian.
Note that the first term in \eqref{H0a1} is not affected by the two-state local phase transformation in \eqref{phasec}, hence the gauge invariant version of \eqref{H0a1} can be written as
\be\label{HLa1}
\hat {\cal H}^{\rm gi}_0 = \frac{\epsilon}{2} \hat \rho_z -  \frac{\Delta}{2}\, \left(|R \rangle \langle L|\, U_{x_R, x_L} + {\rm h. c.}\right)\, ,
\ee
which, in the dipole approximation, reads:
\be\label{HLaa}
\hat {\cal H}^{\rm gi}_0 = \frac{\epsilon}{2} \hat \rho_z  - \frac{\Delta}{2}\, \left(|R \rangle \langle L|\, e^{i q a A} + {\rm h. c.}\right)\, .
\ee
{This} can be expressed as
\be\label{as1}
\hat {\cal H}^{\rm gi}_0 = \frac{\epsilon}{2} \hat \rho_z - \frac{\Delta}{2} \left[ \hat \rho_x \cos{(q a  A)} -  \hat \rho_y \sin{(q a  A)}\right]\,  ,
\ee
which can also be written in the more compact form
\be\label{mcrTLS}
\hat {\cal H}^{\rm gi}_0 = \hat {\cal U} \hat {\cal H}_0 \hat {\cal U}^\dag\, ,
\ee
where
\be\label{U}
\hat {\cal U} = \exp{\left[ i q a A \hat \rho_z /2\right]}\, .
\ee

\begin{figure}[htpb]  
	\centering
	\includegraphics[width= 0.99 \linewidth]{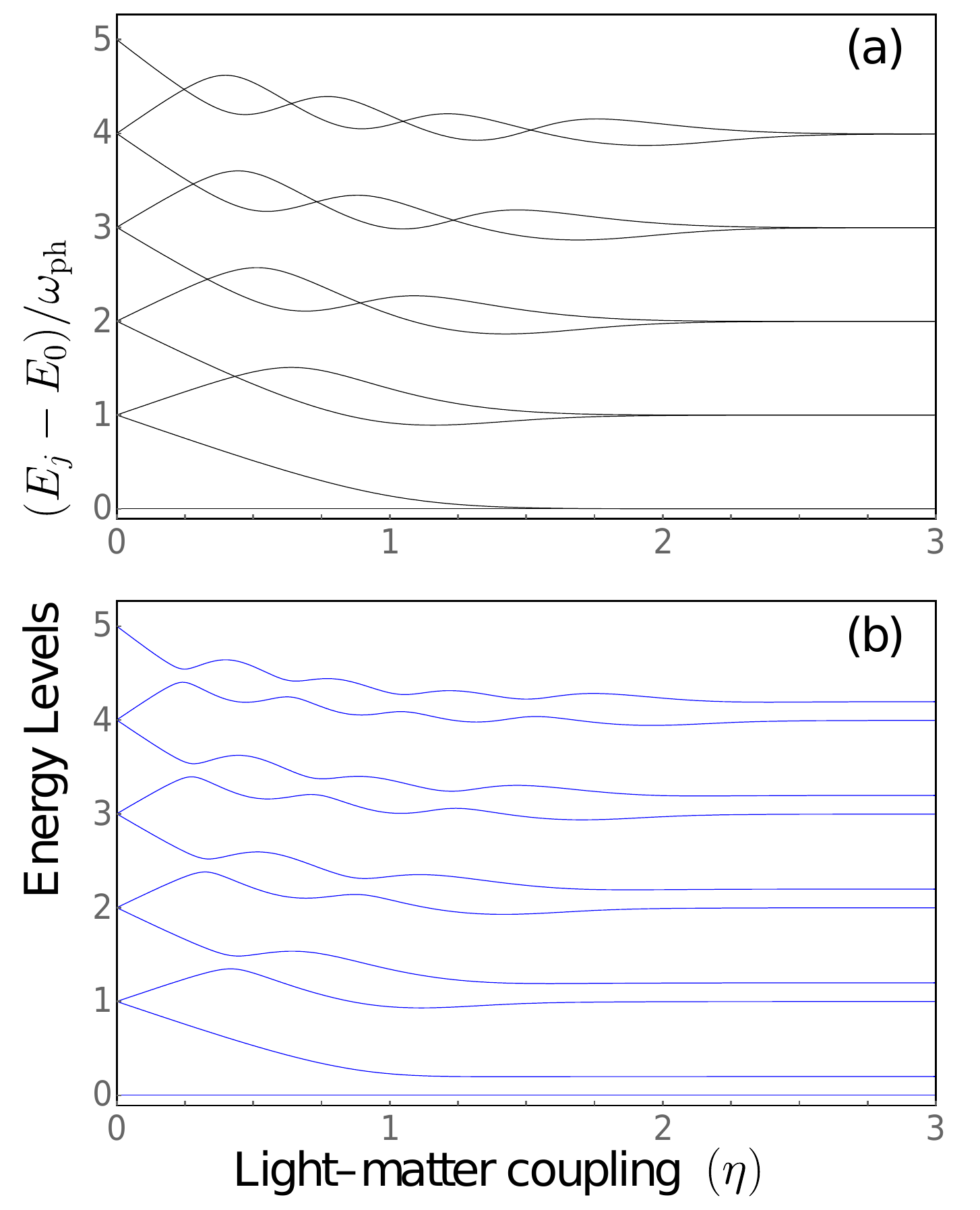}  
	\caption{{\bf Normalized Energy spectra of the QRMs for the symmetric (a) and asymmetric (b) TLS}. Energy  differences  $(E_j - E_0)/\omega_{\rm ph}$ ($E_j$ are energy eigenvalues) as a function of the normalized coupling strength $\eta$, calculated at zero detuning: $\omega_q = \omega_{\rm ph}$. Panel (a) displays the spectrum for the standard QRM (symmetric TLS), while the spectrum for the QRM for the asymmetric TLS is shown in (b).}
	\label{fig3}
\end{figure}
{\em Equations~(\ref{mcrTLS}) and~(\ref{U})
represent the minimal coupling replacement for TLSs, derived directly from the fundamental gauge principle.}

We observe that the operator $\hat {\cal X} = a \hat \rho_z / 2$ represents the geometrical-coordinate operator for the two-state system, with eigenvalues $\pm a/2$.
The Hamiltonian in \eqref{as1} can be directly generalized beyond the dipole approximation with the following replacement:
\be\label{bd}
a A \to \int_{-a/2}^{a/2} \!dx\, A(x)\, .
\ee

Considering a single-mode electromagnetic resonator, the 
total Hamiltonian becomes
\bea\label{33}
\hat {\cal H} &=& \omega_{\rm ph} \hat a^\dag \hat a + \frac{\epsilon}{2} \hat \rho_z  \\ 
&-& \frac{\Delta}{2} \left\{ \hat \rho_x \cos{\left[2 \eta (\hat a + \hat a^\dag)\right]} -  \hat \rho_y \sin{\left[2 \eta (\hat a + \hat a^\dag)\right]}\right\}\, .
\nonumber
\eea

Since the operator $\hat {\cal X}$ is the position operator in the two-state space,  the unitary operator $\hat {\cal U}^\dag = \hat{\cal T}$ also corresponds to the  operator which implements the PZW unitary transformation \cite{Babiker1983}, leading to the dipole-gauge representation,
\bea\label{34}
\hat {\cal H}_d &=& \hat {\cal U}^\dag \hat {\cal H} \hat {\cal U} = \omega_{\rm ph} \hat a^\dag \hat a  + \frac{\epsilon}{2} \hat \rho_z - \frac{\Delta}{2} \hat \rho_x \nonumber\\
&-& i \eta  \omega_{\rm ph} (\hat a - \hat a^\dag) \hat \rho_z +
\eta^2 \hat {\cal I}\, ,
\eea
where we used: $\hat \rho_z^2 = \hat {\cal I}$, with ${\cal I}$  the identity operator for the two-state system.
Note that $\hat {\cal H}_d$ coincides  with the Hamiltonian describing a flux qubit interacting with an $LC$ oscillator \cite{Yoshihara2017}.

Since the Hamiltonians in \eqref{33} and  \eqref{34} are related by a gauge (unitary) transformation, their eigenvalues $E_j$ coincide. Figure~\ref{fig3} displays their energy spectra, defined as $(E_j - E_0)/ \omega_{\rm ph}$ as a function of the normalized coupling strength, where $E_0$ is the ground state energy. The spectra have been obtained at zero detuning: $\omega_q = \omega_{\rm ph}$. In particular, \figref{fig3}(a) displays the energy spectrum in the absence of symmetry breaking ($\epsilon = 0$), namely that of the standard QRM \cite{DiStefano2019}. Panel~\ref{fig3}(b) is obtained using $\epsilon = 0.2 \omega_{\rm ph}$. Such a symmetry breaking gives rise to a number of interesting features. In particular, we observe that the level crossings present in panel~\ref{fig3}(a) convert into avoided-level crossings. The appearance of these splittings is a signature of the hybridization between states with different parity. Note that, in the Jaynes Cummings  model (the QRM after the rotating wave approximation) the number of excitations is conserved. In the QRM, owing to the counter-rotating terms, such a number is no 
longer conserved. However its parity remains a good quantum number \cite{Kockum2018}.
For $\epsilon \neq 0$, also this symmetry is removed.
A peculiar feature of the QRM  consists of energy levels $E_j - E_0$ which tend to become flat and \qum{two-fold degenerate} in the extreme coupling limit. Figure~\ref{fig3}(b) shows that this degeneracy is removed and in the limit $\eta \to \infty$, it is converted into a gap exactly equal to $\epsilon$.

\section{Conclusions}

This work has  discussed the connection between the QRM, a widespread  model in quantum optics, and lattice gauge theory, and shows that the results in Ref.~\cite{DiStefano2019}, obtained with a completely different approach,
{fit} well in the great tradition of lattice gauge theories opened by Kenneth Wilson~\cite{Wiese2013}.
Lattice gauge theories
{constitute a powerful reference example, where it is} possible and also  vital to maintain the gauge invariance of a theory after reducing the infinite amount of information associated to a continuous coordinate \cite{Wiese2013}.

In order to highlight the versatility of the prescription used here, we have presented the gauge invariant formulation in  the case of asymmetric two-state systems interacting with the electromagnetic field, extending 
the results in Ref.~\cite{DiStefano2019} to the case of asymmetric two-state systems interacting with the electromagnetic field. The corresponding energy spectrum, for a single-mode field, as a function of the normalized coupling strength, shows the impact of breaking parity symmetry in the USC regime.
In addition, the method used here allowed us to obtain the gauge-invariant QRM beyond the dipole approximation.

It is our hope that the results and the connection between the QRM and  lattice gauge theory presented here can stimulate the development of lattice gauge models for the study of USC cavity QED in 1D and 2D systems, as well as of interacting electron systems \cite{Savasta1995, Andolina2019, Mordovina2020, Dmytruk2020}. It would also be interesting to apply lattice gauge theory to investigate cavity QED systems beyond the dipole approximation \cite{Andolina2020}.

\vspace{1cm}
{\begin{center}\bf ACKNOWLEDGMENTS \end{center}
	
	F.N. is supported in part by: NTT Research, 
	Army Research Office (ARO) (Grant No. W911NF-18-1-0358), 
	Japan Science and Technology Agency (JST) (via the Q-LEAP program and the CREST Grant No. JPMJCR1676), 
	Japan Society for the Promotion of Science (JSPS) (via the KAKENHI Grant No. JP20H00134 
	and the JSPS-RFBR Grant No. JPJSBP120194828), 
	the Asian Office of Aerospace Research and Development (AOARD), 
	and the Foundational Questions Institute Fund (FQXi) via Grant No. FQXi-IAF19-06.
	S.H. acknowledges funding from 
	the Canadian Foundation for Innovation, and
	the Natural Sciences and Engineering Research Council of Canada.
	%
	S.S. acknowledges the Army Research Office (ARO)
	(Grant No. W911NF1910065).

\bibliography{refs}

\begin{thebibliography}{33}%
\makeatletter
\providecommand \@ifxundefined [1]{%
 \@ifx{#1\undefined}
}%
\providecommand \@ifnum [1]{%
 \ifnum #1\expandafter \@firstoftwo
 \else \expandafter \@secondoftwo
 \fi
}%
\providecommand \@ifx [1]{%
 \ifx #1\expandafter \@firstoftwo
 \else \expandafter \@secondoftwo
 \fi
}%
\providecommand \natexlab [1]{#1}%
\providecommand \enquote  [1]{``#1''}%
\providecommand \bibnamefont  [1]{#1}%
\providecommand \bibfnamefont [1]{#1}%
\providecommand \citenamefont [1]{#1}%
\providecommand \href@noop [0]{\@secondoftwo}%
\providecommand \href [0]{\begingroup \@sanitize@url \@href}%
\providecommand \@href[1]{\@@startlink{#1}\@@href}%
\providecommand \@@href[1]{\endgroup#1\@@endlink}%
\providecommand \@sanitize@url [0]{\catcode `\\12\catcode `\$12\catcode
  `\&12\catcode `\#12\catcode `\^12\catcode `\_12\catcode `\%12\relax}%
\providecommand \@@startlink[1]{}%
\providecommand \@@endlink[0]{}%
\providecommand \url  [0]{\begingroup\@sanitize@url \@url }%
\providecommand \@url [1]{\endgroup\@href {#1}{\urlprefix }}%
\providecommand \urlprefix  [0]{URL }%
\providecommand \Eprint [0]{\href }%
\providecommand \doibase [0]{https://doi.org/}%
\providecommand \selectlanguage [0]{\@gobble}%
\providecommand \bibinfo  [0]{\@secondoftwo}%
\providecommand \bibfield  [0]{\@secondoftwo}%
\providecommand \translation [1]{[#1]}%
\providecommand \BibitemOpen [0]{}%
\providecommand \bibitemStop [0]{}%
\providecommand \bibitemNoStop [0]{.\EOS\space}%
\providecommand \EOS [0]{\spacefactor3000\relax}%
\providecommand \BibitemShut  [1]{\csname bibitem#1\endcsname}%
\let\auto@bib@innerbib\@empty
\bibitem [{\citenamefont {De~Bernardis}\ \emph {et~al.}(2018)\citenamefont
  {De~Bernardis}, \citenamefont {Pilar}, \citenamefont {Jaako}, \citenamefont
  {De~Liberato},\ and\ \citenamefont {Rabl}}]{DeBernardis2018}%
  \BibitemOpen
  \bibfield  {author} {\bibinfo {author} {\bibfnamefont {D.}~\bibnamefont
  {De~Bernardis}}, \bibinfo {author} {\bibfnamefont {P.}~\bibnamefont {Pilar}},
  \bibinfo {author} {\bibfnamefont {T.}~\bibnamefont {Jaako}}, \bibinfo
  {author} {\bibfnamefont {S.}~\bibnamefont {De~Liberato}},\ and\ \bibinfo
  {author} {\bibfnamefont {P.}~\bibnamefont {Rabl}},\ }\bibfield  {title}
  {\bibinfo {title} {Breakdown of gauge invariance in ultrastrong-coupling
  cavity {QED}},\ }\href {https://doi.org/10.1103/PhysRevA.98.053819}
  {\bibfield  {journal} {\bibinfo  {journal} {Phys. Rev. A}\ }\textbf {\bibinfo
  {volume} {98}},\ \bibinfo {pages} {053819} (\bibinfo {year}
  {2018})}\BibitemShut {NoStop}%
\bibitem [{\citenamefont {Stokes}\ and\ \citenamefont
  {Nazir}(2019{\natexlab{a}})}]{Stokes2019}%
  \BibitemOpen
  \bibfield  {author} {\bibinfo {author} {\bibfnamefont {A.}~\bibnamefont
  {Stokes}}\ and\ \bibinfo {author} {\bibfnamefont {A.}~\bibnamefont {Nazir}},\
  }\bibfield  {title} {\bibinfo {title} {Gauge ambiguities imply
  {J}aynes-{C}ummings physics remains valid in ultrastrong coupling {QED}},\
  }\href {https://www.nature.com/articles/s41467-018-08101-0} {\bibfield
  {journal} {\bibinfo  {journal} {Nat. Commun.}\ }\textbf {\bibinfo {volume}
  {10}},\ \bibinfo {pages} {499} (\bibinfo {year}
  {2019}{\natexlab{a}})}\BibitemShut {NoStop}%
\bibitem [{\citenamefont {Stokes}\ and\ \citenamefont
  {Nazir}(2020)}]{Stokes2020a}%
  \BibitemOpen
  \bibfield  {author} {\bibinfo {author} {\bibfnamefont {A.}~\bibnamefont
  {Stokes}}\ and\ \bibinfo {author} {\bibfnamefont {A.}~\bibnamefont {Nazir}},\
  }\bibfield  {title} {\bibinfo {title} {{Gauge non-invariance due to material
  truncation in ultrastrong-coupling QED}},\ }\href
  {https://arxiv.org/abs/2005.06499} {\bibfield  {journal} {\bibinfo  {journal}
  {Preprint at arXiv:2005.06499v1}\ } (\bibinfo {year} {2020})}\BibitemShut
  {NoStop}%
\bibitem [{\citenamefont {Kockum}\ \emph {et~al.}(2019)\citenamefont {Kockum},
  \citenamefont {Miranowicz}, \citenamefont {Liberato}, \citenamefont
  {Savasta},\ and\ \citenamefont {Nori}}]{Kockum2018}%
  \BibitemOpen
  \bibfield  {author} {\bibinfo {author} {\bibfnamefont {A.~F.}\ \bibnamefont
  {Kockum}}, \bibinfo {author} {\bibfnamefont {A.}~\bibnamefont {Miranowicz}},
  \bibinfo {author} {\bibfnamefont {S.~D.}\ \bibnamefont {Liberato}}, \bibinfo
  {author} {\bibfnamefont {S.}~\bibnamefont {Savasta}},\ and\ \bibinfo {author}
  {\bibfnamefont {F.}~\bibnamefont {Nori}},\ }\bibfield  {title} {\bibinfo
  {title} {Ultrastrong coupling between light and matter},\ }\href
  {https://doi.org/10.1038/s42254-018-0006-2} {\bibfield  {journal} {\bibinfo
  {journal} {Nat. Rev. Phys.}\ }\textbf {\bibinfo {volume} {1}},\ \bibinfo
  {pages} {19} (\bibinfo {year} {2019})}\BibitemShut {NoStop}%
\bibitem [{\citenamefont {Forn-D\'{\i}az}\ \emph {et~al.}(2019)\citenamefont
  {Forn-D\'{\i}az}, \citenamefont {Lamata}, \citenamefont {Rico}, \citenamefont
  {Kono},\ and\ \citenamefont {Solano}}]{Forn-Diaz2018}%
  \BibitemOpen
  \bibfield  {author} {\bibinfo {author} {\bibfnamefont {P.}~\bibnamefont
  {Forn-D\'{\i}az}}, \bibinfo {author} {\bibfnamefont {L.}~\bibnamefont
  {Lamata}}, \bibinfo {author} {\bibfnamefont {E.}~\bibnamefont {Rico}},
  \bibinfo {author} {\bibfnamefont {J.}~\bibnamefont {Kono}},\ and\ \bibinfo
  {author} {\bibfnamefont {E.}~\bibnamefont {Solano}},\ }\bibfield  {title}
  {\bibinfo {title} {Ultrastrong coupling regimes of light-matter
  interaction},\ }\href {https://doi.org/10.1103/RevModPhys.91.025005}
  {\bibfield  {journal} {\bibinfo  {journal} {Rev. Mod. Phys.}\ }\textbf
  {\bibinfo {volume} {91}},\ \bibinfo {pages} {025005} (\bibinfo {year}
  {2019})}\BibitemShut {NoStop}%
\bibitem [{\citenamefont {Di~Stefano}\ \emph {et~al.}(2019)\citenamefont
  {Di~Stefano}, \citenamefont {Settineri}, \citenamefont {Macr{\`\i}},
  \citenamefont {Garziano}, \citenamefont {Stassi}, \citenamefont {Savasta},\
  and\ \citenamefont {Nori}}]{DiStefano2019}%
  \BibitemOpen
  \bibfield  {author} {\bibinfo {author} {\bibfnamefont {O.}~\bibnamefont
  {Di~Stefano}}, \bibinfo {author} {\bibfnamefont {A.}~\bibnamefont
  {Settineri}}, \bibinfo {author} {\bibfnamefont {V.}~\bibnamefont
  {Macr{\`\i}}}, \bibinfo {author} {\bibfnamefont {L.}~\bibnamefont
  {Garziano}}, \bibinfo {author} {\bibfnamefont {R.}~\bibnamefont {Stassi}},
  \bibinfo {author} {\bibfnamefont {S.}~\bibnamefont {Savasta}},\ and\ \bibinfo
  {author} {\bibfnamefont {F.}~\bibnamefont {Nori}},\ }\bibfield  {title}
  {\bibinfo {title} {{Resolution of gauge ambiguities in ultrastrong-coupling
  cavity QED}},\ }\href {https://www.nature.com/articles/s41567-019-0534-4}
  {\bibfield  {journal} {\bibinfo  {journal} {Nat. Phys.}\ }\textbf {\bibinfo
  {volume} {15}},\ \bibinfo {pages} {803} (\bibinfo {year} {2019})}\BibitemShut
  {NoStop}%
\bibitem [{\citenamefont {Settineri}\ \emph {et~al.}(2019)\citenamefont
  {Settineri}, \citenamefont {Di~Stefano}, \citenamefont {Zueco}, \citenamefont
  {Hughes}, \citenamefont {Savasta},\ and\ \citenamefont
  {Nori}}]{Settineri2020}%
  \BibitemOpen
  \bibfield  {author} {\bibinfo {author} {\bibfnamefont {A.}~\bibnamefont
  {Settineri}}, \bibinfo {author} {\bibfnamefont {O.}~\bibnamefont
  {Di~Stefano}}, \bibinfo {author} {\bibfnamefont {D.}~\bibnamefont {Zueco}},
  \bibinfo {author} {\bibfnamefont {S.}~\bibnamefont {Hughes}}, \bibinfo
  {author} {\bibfnamefont {S.}~\bibnamefont {Savasta}},\ and\ \bibinfo {author}
  {\bibfnamefont {F.}~\bibnamefont {Nori}},\ }\bibfield  {title} {\bibinfo
  {title} {Gauge freedom, quantum measurements, and time-dependent interactions
  in cavity and circuit {QED}},\ }\href {https://arxiv.org/abs/1912.08548}
  {\bibfield  {journal} {\bibinfo  {journal} {Preprint at arXiv: 1912.08548}\ }
  (\bibinfo {year} {2019})}\BibitemShut {NoStop}%
\bibitem [{\citenamefont {Garziano}\ \emph {et~al.}(2020)\citenamefont
  {Garziano}, \citenamefont {Settineri}, \citenamefont {Di~Stefano},
  \citenamefont {Savasta},\ and\ \citenamefont {Nori}}]{Garziano2020}%
  \BibitemOpen
  \bibfield  {author} {\bibinfo {author} {\bibfnamefont {L.}~\bibnamefont
  {Garziano}}, \bibinfo {author} {\bibfnamefont {A.}~\bibnamefont {Settineri}},
  \bibinfo {author} {\bibfnamefont {O.}~\bibnamefont {Di~Stefano}}, \bibinfo
  {author} {\bibfnamefont {S.}~\bibnamefont {Savasta}},\ and\ \bibinfo {author}
  {\bibfnamefont {F.}~\bibnamefont {Nori}},\ }\bibfield  {title} {\bibinfo
  {title} {Gauge invariance of the dicke and hopfield models},\ }\href
  {https://doi.org/10.1103/PhysRevA.102.023718} {\bibfield  {journal} {\bibinfo
   {journal} {Phys. Rev. A}\ }\textbf {\bibinfo {volume} {102}},\ \bibinfo
  {pages} {023718} (\bibinfo {year} {2020})}\BibitemShut {NoStop}%
\bibitem [{\citenamefont {Savasta}\ \emph {et~al.}(2020)\citenamefont
  {Savasta}, \citenamefont {Stefano},\ and\ \citenamefont {Nori}}]{SDN2020}%
  \BibitemOpen
  \bibfield  {author} {\bibinfo {author} {\bibfnamefont {S.}~\bibnamefont
  {Savasta}}, \bibinfo {author} {\bibfnamefont {O.~D.}\ \bibnamefont
  {Stefano}},\ and\ \bibinfo {author} {\bibfnamefont {F.}~\bibnamefont
  {Nori}},\ }\bibfield  {title} {\bibinfo {title} {{TRK} sum rule for
  interacting photons},\ }\href {https://arxiv.org/abs/2002.02139} {\bibfield
  {journal} {\bibinfo  {journal} {{\em to appear on} Nanophotonics, Preprint at
  arXiv:2002.02139}\ } (\bibinfo {year} {2020})}\BibitemShut {NoStop}%
\bibitem [{\citenamefont {Le~Boit{\'e}}(2020)}]{LeBoite2020}%
  \BibitemOpen
  \bibfield  {author} {\bibinfo {author} {\bibfnamefont {A.}~\bibnamefont
  {Le~Boit{\'e}}},\ }\bibfield  {title} {\bibinfo {title} {Theoretical methods
  for ultrastrong light--matter interactions},\ }\href
  {https://onlinelibrary.wiley.com/doi/full/10.1002/qute.201900140} {\bibfield
  {journal} {\bibinfo  {journal} {Adv. Quantum Technol.}\ ,\ \bibinfo {pages}
  {1900140}} (\bibinfo {year} {2020})}\BibitemShut {NoStop}%
\bibitem [{\citenamefont {Wiese}(2013)}]{Wiese2013}%
  \BibitemOpen
  \bibfield  {author} {\bibinfo {author} {\bibfnamefont {U.-J.}\ \bibnamefont
  {Wiese}},\ }\bibfield  {title} {\bibinfo {title} {Ultracold quantum gases and
  lattice systems: quantum simulation of lattice gauge theories},\ }\href
  {https://onlinelibrary.wiley.com/doi/full/10.1002/andp.201300104} {\bibfield
  {journal} {\bibinfo  {journal} {Ann. Phys.}\ }\textbf {\bibinfo {volume}
  {525}},\ \bibinfo {pages} {777} (\bibinfo {year} {2013})}\BibitemShut
  {NoStop}%
\bibitem [{\citenamefont {Peierls}(1933)}]{Peirls1933}%
  \BibitemOpen
  \bibfield  {author} {\bibinfo {author} {\bibfnamefont {R.}~\bibnamefont
  {Peierls}},\ }\href@noop {} {\bibfield  {journal} {\bibinfo  {journal} {Z.
  Phys.}\ }\textbf {\bibinfo {volume} {80}},\ \bibinfo {pages} {763} (\bibinfo
  {year} {1933})}\BibitemShut {NoStop}%
\bibitem [{\citenamefont {Luttinger}(1951)}]{Luttinger1951}%
  \BibitemOpen
  \bibfield  {author} {\bibinfo {author} {\bibfnamefont {J.~M.}\ \bibnamefont
  {Luttinger}},\ }\bibfield  {title} {\bibinfo {title} {The effect of a
  magnetic field on electrons in a periodic potential},\ }\href
  {https://doi.org/10.1103/PhysRev.84.814} {\bibfield  {journal} {\bibinfo
  {journal} {Phys. Rev.}\ }\textbf {\bibinfo {volume} {84}},\ \bibinfo {pages}
  {814} (\bibinfo {year} {1951})}\BibitemShut {NoStop}%
\bibitem [{\citenamefont {Hofstadter}(1976)}]{Hofstadter1976}%
  \BibitemOpen
  \bibfield  {author} {\bibinfo {author} {\bibfnamefont {D.~R.}\ \bibnamefont
  {Hofstadter}},\ }\bibfield  {title} {\bibinfo {title} {{Energy levels and
  wave functions of Bloch electrons in rational and irrational magnetic
  fields}},\ }\href {https://doi.org/10.1103/PhysRevB.14.2239} {\bibfield
  {journal} {\bibinfo  {journal} {Phys. Rev. B}\ }\textbf {\bibinfo {volume}
  {14}},\ \bibinfo {pages} {2239} (\bibinfo {year} {1976})}\BibitemShut
  {NoStop}%
\bibitem [{\citenamefont {Graf}\ and\ \citenamefont {Vogl}(1995)}]{Graf1995}%
  \BibitemOpen
  \bibfield  {author} {\bibinfo {author} {\bibfnamefont {M.}~\bibnamefont
  {Graf}}\ and\ \bibinfo {author} {\bibfnamefont {P.}~\bibnamefont {Vogl}},\
  }\bibfield  {title} {\bibinfo {title} {Electromagnetic fields and dielectric
  response in empirical tight-binding theory},\ }\href
  {https://doi.org/10.1103/PhysRevB.51.4940} {\bibfield  {journal} {\bibinfo
  {journal} {Phys. Rev. B}\ }\textbf {\bibinfo {volume} {51}},\ \bibinfo
  {pages} {4940} (\bibinfo {year} {1995})}\BibitemShut {NoStop}%
\bibitem [{\citenamefont {Maggiore}(2005)}]{Maggiore2005}%
  \BibitemOpen
  \bibfield  {author} {\bibinfo {author} {\bibfnamefont {M.}~\bibnamefont
  {Maggiore}},\ }\href@noop {} {\emph {\bibinfo {title} {A modern introduction
  to quantum field theory}}},\ \bibinfo {series} {Oxford Series in Physics}\
  No.~\bibinfo {number} {12}\ (\bibinfo  {publisher} {Oxford {U}niversity
  {P}ress},\ \bibinfo {year} {2005})\BibitemShut {NoStop}%
\bibitem [{\citenamefont {Leggett}\ \emph {et~al.}(1987)\citenamefont
  {Leggett}, \citenamefont {Chakravarty}, \citenamefont {Dorsey}, \citenamefont
  {Fisher}, \citenamefont {Garg},\ and\ \citenamefont {Zwerger}}]{Leggett1987}%
  \BibitemOpen
  \bibfield  {author} {\bibinfo {author} {\bibfnamefont {A.~J.}\ \bibnamefont
  {Leggett}}, \bibinfo {author} {\bibfnamefont {S.}~\bibnamefont
  {Chakravarty}}, \bibinfo {author} {\bibfnamefont {A.~T.}\ \bibnamefont
  {Dorsey}}, \bibinfo {author} {\bibfnamefont {M.~P.~A.}\ \bibnamefont
  {Fisher}}, \bibinfo {author} {\bibfnamefont {A.}~\bibnamefont {Garg}},\ and\
  \bibinfo {author} {\bibfnamefont {W.}~\bibnamefont {Zwerger}},\ }\bibfield
  {title} {\bibinfo {title} {Dynamics of the dissipative two-state system},\
  }\href {https://doi.org/10.1103/RevModPhys.59.1} {\bibfield  {journal}
  {\bibinfo  {journal} {Rev. Mod. Phys.}\ }\textbf {\bibinfo {volume} {59}},\
  \bibinfo {pages} {1} (\bibinfo {year} {1987})}\BibitemShut {NoStop}%
\bibitem [{\citenamefont {Stokes}\ and\ \citenamefont
  {Nazir}(2019{\natexlab{b}})}]{Stokes2019a}%
  \BibitemOpen
  \bibfield  {author} {\bibinfo {author} {\bibfnamefont {A.}~\bibnamefont
  {Stokes}}\ and\ \bibinfo {author} {\bibfnamefont {A.}~\bibnamefont {Nazir}},\
  }\bibfield  {title} {\bibinfo {title} {Ultrastrong time-dependent
  light-matter interactions are gauge-relative},\ }\href
  {https://arxiv.org/abs/1902.05160} {\bibfield  {journal} {\bibinfo  {journal}
  {Preprint at arXiv:1902.05160}\ } (\bibinfo {year}
  {2019}{\natexlab{b}})}\BibitemShut {NoStop}%
\bibitem [{\citenamefont {Wilson}(1974)}]{Wilson1974}%
  \BibitemOpen
  \bibfield  {author} {\bibinfo {author} {\bibfnamefont {K.~G.}\ \bibnamefont
  {Wilson}},\ }\bibfield  {title} {\bibinfo {title} {Confinement of quarks},\
  }\href {https://doi.org/10.1103/PhysRevD.10.2445} {\bibfield  {journal}
  {\bibinfo  {journal} {Phys. Rev. D}\ }\textbf {\bibinfo {volume} {10}},\
  \bibinfo {pages} {2445} (\bibinfo {year} {1974})}\BibitemShut {NoStop}%
\bibitem [{\citenamefont {Lang}(2010)}]{Lang2010}%
  \BibitemOpen
  \bibfield  {author} {\bibinfo {author} {\bibfnamefont {C.~B.}\ \bibnamefont
  {Lang}},\ }\href@noop {} {\emph {\bibinfo {title} {Quantum chromodynamics on
  the lattice: an introductory presentation}}}\ (\bibinfo  {publisher}
  {Springer},\ \bibinfo {year} {2010})\BibitemShut {NoStop}%
\bibitem [{\citenamefont {Niemczyk}\ \emph {et~al.}(2010)\citenamefont
  {Niemczyk}, \citenamefont {Deppe}, \citenamefont {Huebl}, \citenamefont
  {Menzel}, \citenamefont {Hocke}, \citenamefont {Schwarz}, \citenamefont
  {Garcia-Ripoll}, \citenamefont {Zueco}, \citenamefont {H{\"{u}}mmer},
  \citenamefont {Solano}, \citenamefont {Marx},\ and\ \citenamefont
  {Gross}}]{Niemczyk2010}%
  \BibitemOpen
  \bibfield  {author} {\bibinfo {author} {\bibfnamefont {T.}~\bibnamefont
  {Niemczyk}}, \bibinfo {author} {\bibfnamefont {F.}~\bibnamefont {Deppe}},
  \bibinfo {author} {\bibfnamefont {H.}~\bibnamefont {Huebl}}, \bibinfo
  {author} {\bibfnamefont {E.~P.}\ \bibnamefont {Menzel}}, \bibinfo {author}
  {\bibfnamefont {F.}~\bibnamefont {Hocke}}, \bibinfo {author} {\bibfnamefont
  {M.~J.}\ \bibnamefont {Schwarz}}, \bibinfo {author} {\bibfnamefont {J.~J.}\
  \bibnamefont {Garcia-Ripoll}}, \bibinfo {author} {\bibfnamefont
  {D.}~\bibnamefont {Zueco}}, \bibinfo {author} {\bibfnamefont
  {T.}~\bibnamefont {H{\"{u}}mmer}}, \bibinfo {author} {\bibfnamefont
  {E.}~\bibnamefont {Solano}}, \bibinfo {author} {\bibfnamefont
  {A.}~\bibnamefont {Marx}},\ and\ \bibinfo {author} {\bibfnamefont
  {R.}~\bibnamefont {Gross}},\ }\bibfield  {title} {\bibinfo {title} {{Circuit
  quantum electrodynamics in the ultrastrong-coupling regime}},\ }\href
  {https://doi.org/10.1038/nphys1730} {\bibfield  {journal} {\bibinfo
  {journal} {Nat. Phys.}\ }\textbf {\bibinfo {volume} {6}},\ \bibinfo {pages}
  {772} (\bibinfo {year} {2010})}\BibitemShut {NoStop}%
\bibitem [{\citenamefont {Ridolfo}\ \emph {et~al.}(2012)\citenamefont
  {Ridolfo}, \citenamefont {Leib}, \citenamefont {Savasta},\ and\ \citenamefont
  {Hartmann}}]{Ridolfo2012}%
  \BibitemOpen
  \bibfield  {author} {\bibinfo {author} {\bibfnamefont {A.}~\bibnamefont
  {Ridolfo}}, \bibinfo {author} {\bibfnamefont {M.}~\bibnamefont {Leib}},
  \bibinfo {author} {\bibfnamefont {S.}~\bibnamefont {Savasta}},\ and\ \bibinfo
  {author} {\bibfnamefont {M.~J.}\ \bibnamefont {Hartmann}},\ }\bibfield
  {title} {\bibinfo {title} {Photon blockade in the ultrastrong coupling
  regime},\ }\href {https://doi.org/10.1103/PhysRevLett.109.193602} {\bibfield
  {journal} {\bibinfo  {journal} {Phys. Rev. Lett.}\ }\textbf {\bibinfo
  {volume} {109}},\ \bibinfo {pages} {193602} (\bibinfo {year}
  {2012})}\BibitemShut {NoStop}%
\bibitem [{\citenamefont {Garziano}\ \emph {et~al.}(2015)\citenamefont
  {Garziano}, \citenamefont {Stassi}, \citenamefont {Macr{\`{i}}},
  \citenamefont {Kockum}, \citenamefont {Savasta},\ and\ \citenamefont
  {Nori}}]{Garziano2015}%
  \BibitemOpen
  \bibfield  {author} {\bibinfo {author} {\bibfnamefont {L.}~\bibnamefont
  {Garziano}}, \bibinfo {author} {\bibfnamefont {R.}~\bibnamefont {Stassi}},
  \bibinfo {author} {\bibfnamefont {V.}~\bibnamefont {Macr{\`{i}}}}, \bibinfo
  {author} {\bibfnamefont {A.~F.}\ \bibnamefont {Kockum}}, \bibinfo {author}
  {\bibfnamefont {S.}~\bibnamefont {Savasta}},\ and\ \bibinfo {author}
  {\bibfnamefont {F.}~\bibnamefont {Nori}},\ }\bibfield  {title} {\bibinfo
  {title} {{Multiphoton quantum Rabi oscillations in ultrastrong cavity QED}},\
  }\href {https://doi.org/10.1103/PhysRevA.92.063830} {\bibfield  {journal}
  {\bibinfo  {journal} {Phys. Rev. A}\ }\textbf {\bibinfo {volume} {92}},\
  \bibinfo {pages} {063830} (\bibinfo {year} {2015})}\BibitemShut {NoStop}%
\bibitem [{\citenamefont {Garziano}\ \emph {et~al.}(2016)\citenamefont
  {Garziano}, \citenamefont {Macr{\`{i}}}, \citenamefont {Stassi},
  \citenamefont {{Di Stefano}}, \citenamefont {Nori},\ and\ \citenamefont
  {Savasta}}]{Garziano2016}%
  \BibitemOpen
  \bibfield  {author} {\bibinfo {author} {\bibfnamefont {L.}~\bibnamefont
  {Garziano}}, \bibinfo {author} {\bibfnamefont {V.}~\bibnamefont
  {Macr{\`{i}}}}, \bibinfo {author} {\bibfnamefont {R.}~\bibnamefont {Stassi}},
  \bibinfo {author} {\bibfnamefont {O.}~\bibnamefont {{Di Stefano}}}, \bibinfo
  {author} {\bibfnamefont {F.}~\bibnamefont {Nori}},\ and\ \bibinfo {author}
  {\bibfnamefont {S.}~\bibnamefont {Savasta}},\ }\bibfield  {title} {\bibinfo
  {title} {One photon can simultaneously excite two or more atoms},\ }\href
  {https://doi.org/10.1103/PhysRevLett.117.043601} {\bibfield  {journal}
  {\bibinfo  {journal} {Phys. Rev. Lett.}\ }\textbf {\bibinfo {volume} {117}},\
  \bibinfo {pages} {043601} (\bibinfo {year} {2016})}\BibitemShut {NoStop}%
\bibitem [{\citenamefont {Yoshihara}\ \emph {et~al.}(2017)\citenamefont
  {Yoshihara}, \citenamefont {Fuse}, \citenamefont {Ashhab}, \citenamefont
  {Kakuyanagi}, \citenamefont {Saito},\ and\ \citenamefont
  {Semba}}]{Yoshihara2017}%
  \BibitemOpen
  \bibfield  {author} {\bibinfo {author} {\bibfnamefont {F.}~\bibnamefont
  {Yoshihara}}, \bibinfo {author} {\bibfnamefont {T.}~\bibnamefont {Fuse}},
  \bibinfo {author} {\bibfnamefont {S.}~\bibnamefont {Ashhab}}, \bibinfo
  {author} {\bibfnamefont {K.}~\bibnamefont {Kakuyanagi}}, \bibinfo {author}
  {\bibfnamefont {S.}~\bibnamefont {Saito}},\ and\ \bibinfo {author}
  {\bibfnamefont {K.}~\bibnamefont {Semba}},\ }\bibfield  {title} {\bibinfo
  {title} {{Superconducting qubit-oscillator circuit beyond the
  ultrastrong-coupling regime}},\ }\href {https://doi.org/10.1038/nphys3906}
  {\bibfield  {journal} {\bibinfo  {journal} {Nat. Phys.}\ }\textbf {\bibinfo
  {volume} {13}},\ \bibinfo {pages} {44} (\bibinfo {year} {2017})}\BibitemShut
  {NoStop}%
\bibitem [{\citenamefont {Kockum}\ \emph {et~al.}(2017)\citenamefont {Kockum},
  \citenamefont {Miranowicz}, \citenamefont {Macr{\`{i}}}, \citenamefont
  {Savasta},\ and\ \citenamefont {Nori}}]{Kockum2017a}%
  \BibitemOpen
  \bibfield  {author} {\bibinfo {author} {\bibfnamefont {A.~F.}\ \bibnamefont
  {Kockum}}, \bibinfo {author} {\bibfnamefont {A.}~\bibnamefont {Miranowicz}},
  \bibinfo {author} {\bibfnamefont {V.}~\bibnamefont {Macr{\`{i}}}}, \bibinfo
  {author} {\bibfnamefont {S.}~\bibnamefont {Savasta}},\ and\ \bibinfo {author}
  {\bibfnamefont {F.}~\bibnamefont {Nori}},\ }\bibfield  {title} {\bibinfo
  {title} {{Deterministic quantum nonlinear optics with single atoms and
  virtual photons}},\ }\href {https://doi.org/10.1103/PhysRevA.95.063849}
  {\bibfield  {journal} {\bibinfo  {journal} {Phys. Rev. A}\ }\textbf {\bibinfo
  {volume} {95}},\ \bibinfo {pages} {063849} (\bibinfo {year}
  {2017})}\BibitemShut {NoStop}%
\bibitem [{\citenamefont {Stassi}\ \emph {et~al.}(2017)\citenamefont {Stassi},
  \citenamefont {Macr{\`{i}}}, \citenamefont {Kockum}, \citenamefont {{Di
  Stefano}}, \citenamefont {Miranowicz}, \citenamefont {Savasta},\ and\
  \citenamefont {Nori}}]{Stassi2017}%
  \BibitemOpen
  \bibfield  {author} {\bibinfo {author} {\bibfnamefont {R.}~\bibnamefont
  {Stassi}}, \bibinfo {author} {\bibfnamefont {V.}~\bibnamefont {Macr{\`{i}}}},
  \bibinfo {author} {\bibfnamefont {A.~F.}\ \bibnamefont {Kockum}}, \bibinfo
  {author} {\bibfnamefont {O.}~\bibnamefont {{Di Stefano}}}, \bibinfo {author}
  {\bibfnamefont {A.}~\bibnamefont {Miranowicz}}, \bibinfo {author}
  {\bibfnamefont {S.}~\bibnamefont {Savasta}},\ and\ \bibinfo {author}
  {\bibfnamefont {F.}~\bibnamefont {Nori}},\ }\bibfield  {title} {\bibinfo
  {title} {{Quantum nonlinear optics without photons}},\ }\href
  {https://doi.org/10.1103/PhysRevA.96.023818} {\bibfield  {journal} {\bibinfo
  {journal} {Phys. Rev. A}\ }\textbf {\bibinfo {volume} {96}},\ \bibinfo
  {pages} {023818} (\bibinfo {year} {2017})}\BibitemShut {NoStop}%
\bibitem [{\citenamefont {Babiker}\ and\ \citenamefont
  {Loudon}(1983)}]{Babiker1983}%
  \BibitemOpen
  \bibfield  {author} {\bibinfo {author} {\bibfnamefont {M.}~\bibnamefont
  {Babiker}}\ and\ \bibinfo {author} {\bibfnamefont {R.}~\bibnamefont
  {Loudon}},\ }\bibfield  {title} {\bibinfo {title} {Derivation of the
  {P}ower-{Z}ienau-{W}oolley {H}amiltonian in quantum electrodynamics by gauge
  transformation},\ }\href {https://doi.org/10.1098/rspa.1983.0022} {\bibfield
  {journal} {\bibinfo  {journal} {Proc. R. Soc. Lond. A}\ }\textbf {\bibinfo
  {volume} {385}},\ \bibinfo {pages} {439} (\bibinfo {year}
  {1983})}\BibitemShut {NoStop}%
\bibitem [{\citenamefont {Savasta}\ and\ \citenamefont
  {Girlanda}(1995)}]{Savasta1995}%
  \BibitemOpen
  \bibfield  {author} {\bibinfo {author} {\bibfnamefont {S.}~\bibnamefont
  {Savasta}}\ and\ \bibinfo {author} {\bibfnamefont {R.}~\bibnamefont
  {Girlanda}},\ }\bibfield  {title} {\bibinfo {title} {The particle-photon
  interaction in systems descrided by model {H}amiltonians in second
  quantization},\ }\href
  {https://www.sciencedirect.com/science/article/pii/0038109895002421}
  {\bibfield  {journal} {\bibinfo  {journal} {Solid State Commun.}\ }\textbf
  {\bibinfo {volume} {96}},\ \bibinfo {pages} {517} (\bibinfo {year}
  {1995})}\BibitemShut {NoStop}%
\bibitem [{\citenamefont {Andolina}\ \emph {et~al.}(2019)\citenamefont
  {Andolina}, \citenamefont {Pellegrino}, \citenamefont {Giovannetti},
  \citenamefont {MacDonald},\ and\ \citenamefont {Polini}}]{Andolina2019}%
  \BibitemOpen
  \bibfield  {author} {\bibinfo {author} {\bibfnamefont {G.~M.}\ \bibnamefont
  {Andolina}}, \bibinfo {author} {\bibfnamefont {F.~M.~D.}\ \bibnamefont
  {Pellegrino}}, \bibinfo {author} {\bibfnamefont {V.}~\bibnamefont
  {Giovannetti}}, \bibinfo {author} {\bibfnamefont {A.~H.}\ \bibnamefont
  {MacDonald}},\ and\ \bibinfo {author} {\bibfnamefont {M.}~\bibnamefont
  {Polini}},\ }\bibfield  {title} {\bibinfo {title} {Cavity quantum
  electrodynamics of strongly correlated electron systems: A no-go theorem for
  photon condensation},\ }\href {https://doi.org/10.1103/PhysRevB.100.121109}
  {\bibfield  {journal} {\bibinfo  {journal} {Phys. Rev. B}\ }\textbf {\bibinfo
  {volume} {100}},\ \bibinfo {pages} {121109} (\bibinfo {year}
  {2019})}\BibitemShut {NoStop}%
\bibitem [{\citenamefont {Mordovina}\ \emph {et~al.}(2020)\citenamefont
  {Mordovina}, \citenamefont {Bungey}, \citenamefont {Appel}, \citenamefont
  {Knowles}, \citenamefont {Rubio},\ and\ \citenamefont
  {Manby}}]{Mordovina2020}%
  \BibitemOpen
  \bibfield  {author} {\bibinfo {author} {\bibfnamefont {U.}~\bibnamefont
  {Mordovina}}, \bibinfo {author} {\bibfnamefont {C.}~\bibnamefont {Bungey}},
  \bibinfo {author} {\bibfnamefont {H.}~\bibnamefont {Appel}}, \bibinfo
  {author} {\bibfnamefont {P.~J.}\ \bibnamefont {Knowles}}, \bibinfo {author}
  {\bibfnamefont {A.}~\bibnamefont {Rubio}},\ and\ \bibinfo {author}
  {\bibfnamefont {F.~R.}\ \bibnamefont {Manby}},\ }\bibfield  {title} {\bibinfo
  {title} {Polaritonic coupled-cluster theory},\ }\href
  {https://doi.org/10.1103/PhysRevResearch.2.023262} {\bibfield  {journal}
  {\bibinfo  {journal} {Phys. Rev. Research}\ }\textbf {\bibinfo {volume}
  {2}},\ \bibinfo {pages} {023262} (\bibinfo {year} {2020})}\BibitemShut
  {NoStop}%
\bibitem [{\citenamefont {Dmytruk}\ and\ \citenamefont
  {Schir\'{o}}(2020)}]{Dmytruk2020}%
  \BibitemOpen
  \bibfield  {author} {\bibinfo {author} {\bibfnamefont {O.}~\bibnamefont
  {Dmytruk}}\ and\ \bibinfo {author} {\bibfnamefont {M.}~\bibnamefont
  {Schir\'{o}}},\ }\bibfield  {title} {\bibinfo {title} {Gauge fixing for
  strongly correlated electrons coupled to quantum light},\ }\href
  {https://arxiv.org/abs/2009.11088} {\bibfield  {journal} {\bibinfo  {journal}
  {Preprint at arXiv:1902.05160}\ } (\bibinfo {year} {2020})}\BibitemShut
  {NoStop}%
\bibitem [{\citenamefont {Andolina}\ \emph {et~al.}(2020)\citenamefont
  {Andolina}, \citenamefont {Pellegrino}, \citenamefont {Giovannetti},
  \citenamefont {MacDonald},\ and\ \citenamefont {Polini}}]{Andolina2020}%
  \BibitemOpen
  \bibfield  {author} {\bibinfo {author} {\bibfnamefont {G.~M.}\ \bibnamefont
  {Andolina}}, \bibinfo {author} {\bibfnamefont {F.~M.~D.}\ \bibnamefont
  {Pellegrino}}, \bibinfo {author} {\bibfnamefont {V.}~\bibnamefont
  {Giovannetti}}, \bibinfo {author} {\bibfnamefont {A.~H.}\ \bibnamefont
  {MacDonald}},\ and\ \bibinfo {author} {\bibfnamefont {M.}~\bibnamefont
  {Polini}},\ }\bibfield  {title} {\bibinfo {title} {Theory of photon
  condensation in a spatially varying electromagnetic field},\ }\href
  {https://doi.org/10.1103/PhysRevB.102.125137} {\bibfield  {journal} {\bibinfo
   {journal} {Phys. Rev. B}\ }\textbf {\bibinfo {volume} {102}},\ \bibinfo
  {pages} {125137} (\bibinfo {year} {2020})}\BibitemShut {NoStop}%
\end{thebibliography}%

\end{document}